\documentclass{vgtc}                          

\ifpdf
  \pdfoutput=1\relax                   
  \usepackage{graphicx}                
  \DeclareGraphicsExtensions{.pdf,.png,.jpg,.jpeg} 
\else
  \usepackage{graphicx}                
  \DeclareGraphicsExtensions{.eps}     
\fi%

\graphicspath{{figures/}{pictures/}{images/}{./}} 

\usepackage{microtype}                 
\PassOptionsToPackage{warn}{textcomp}  
\usepackage{textcomp}                  
\usepackage{mathptmx}                  
\usepackage{times}                     
\usepackage{cite}                      
\usepackage{enumitem}                  
\usepackage[dvipsnames]{xcolor}                    

\definecolor{bio-indigo}{RGB}{65, 86, 161}

\onlineid{0}

\vgtccategory{Short}

\vgtcinsertpkg

\title{A Review of Geospatial Content in IEEE Visualization Publications}

\author{Alexander Yoshizumi\thanks{e-mail: ayoshiz@ncsu.edu || lgtateos@ncsu.edu} %
\and Megan M. Coffer\thanks{These authors share second authorship and are listed alphabetically.} %
\and Elyssa L. Collins\footnotemark[2]
\and Mollie D. Gaines\footnotemark[2] 
\and Xiaojie Gao\footnotemark[2]
\and Kate Jones\footnotemark[2] 
\and Ian R. McGregor\footnotemark[2]
\and Katie A. McQuillan\footnotemark[2] 
\and Vinicius Perin\footnotemark[2] 
\and Laura M. Tomkins\footnotemark[2]
\and Thom Worm\footnotemark[2] 
\and Laura Tateosian\footnotemark[1]
}
\affiliation{\scriptsize Center for Geospatial Analytics, North Carolina State University}

\teaser{
  \centering
  \includegraphics[width=6.7in]{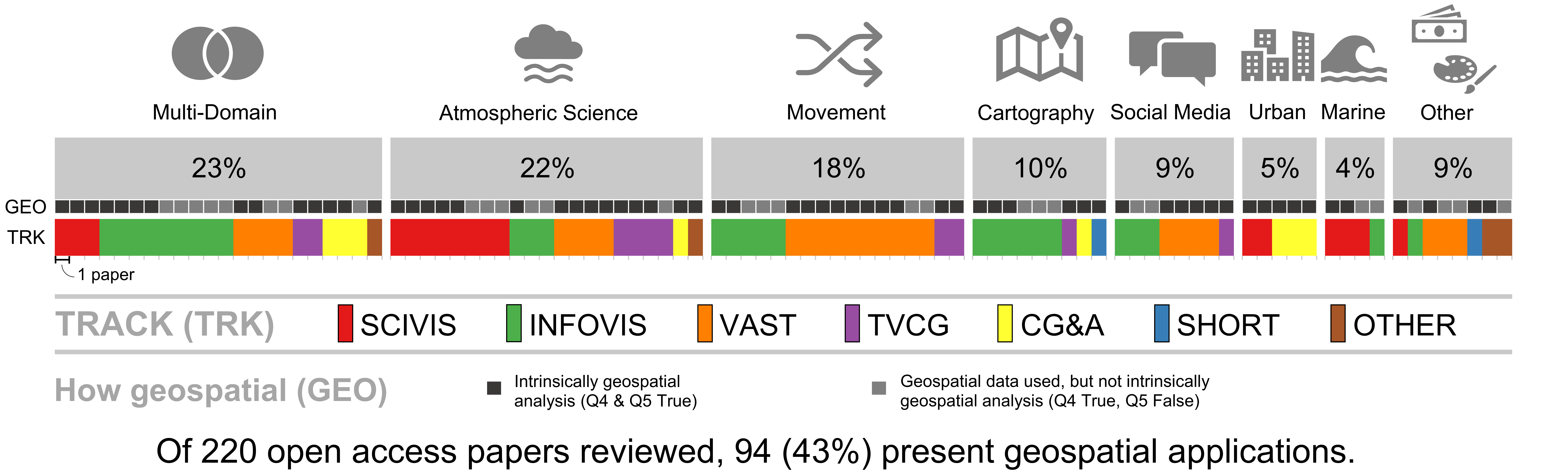}
  \caption{
 Attributes of \textcolor{black}{94} IEEE VIS papers from years 2017-2019 found to have geospatial content.   From top to bottom: \textcolor{black}{data domain}, geospatial nature of the paper \textcolor{black}{(GEO)}, and VIS Conference paper type and track \textcolor{black}{(TRK)} are shown for each paper.  Percentages on the top band (lightest gray bars) correspond to \textcolor{black}{data domain} types. The \textcolor{black}{GEO} band marks papers as either \textcolor{black}{containing both geospatial data and a geospatial analysis (dark gray) or geospatial data only (light gray)}. The \textcolor{black}{TRK} band is colored by the VIS Conference paper types and tracks listed on Open Access VIS~\cite{Har:2018}.} 
  \label{fig:teaser}
}

\abstract{Geospatial analysis is crucial for addressing many of the world's most pressing challenges. Given this, there is immense value in improving and expanding the visualization techniques used to communicate geospatial data. In this work, we explore this important intersection -- between geospatial analytics and visualization -- by examining a set of recent  IEEE VIS Conference papers (a selection from 2017-2019) to assess the inclusion of geospatial data and geospatial analyses within these papers.   After removing the papers with no geospatial data, we organize the remaining literature \textcolor{black}{into geospatial data domain categories} and provide insight into how these categories relate to VIS Conference paper types. We also contextualize our results by investigating the use of geospatial terms in IEEE Visualization publications over the last 30 years. Our work provides an understanding of the quantity and role of geospatial subject matter in recent IEEE VIS publications and supplies a foundation for future meta-analytical work around geospatial analytics and geovisualization that may shed light on opportunities for innovation.%
} 

\CCScatlist{
    \CCScatTwelve{Human-centered computing}{Visu\-al\-iza\-tion}{Visualization application domains}{Geographic visualization}
}

\begin{document}

\maketitle



\section{Introduction}
Geospatial factors play a key role in many of the world's most pressing challenges -- pandemics, plant pest and pathogen spread, natural disasters, urban sprawl, pollution, human trafficking, food scarcity, and transportation (to name just a few). Additionally, as technologies such as GPS-equipped mobile devices, remote sensing satellites, and drones have proliferated, the centrality of georeferenced data has only continued to grow. The complexity and volume of the data and the importance of the issues at stake drive a need for innovative visualization tools to support exploration and communication of geospatial information.

As a focal event for the visualization community, the IEEE Visualization (VIS) Conference profoundly influences the agenda for research in the visualization space. This influence includes identifying new research directions, investigating novel analyses, and presenting results that support a wide array of disciplines, geospatial analytics included. Given this, understanding how geospatial subject matter is covered within the context of VIS conferences is important, as it can shed light on the intersection of geospatial analytics and visualization, highlighting areas of interest and revealing opportunities for innovation.

In this work, we present our efforts to begin to unravel these connections. With a combination of word searches and close readings, we examine the role that geospatial subject matter plays in the IEEE visualization research space, and we provide some fundamental context for how geospatial subject matter has been used within IEEE VIS Conference publications. Specifically, we offer the following contributions:
\begin{itemize}[noitemsep]
    \item{Descriptive analysis of the use of geospatial subject matter in IEEE VIS Conference papers from 2017 to 2019.}
    \item{Temporal contextualization regarding how the role of geospatial subject matter varied across years.}
    \item{\textcolor{black}{Categorization of the 2017 to 2019 papers that leveraged geospatial data by data domain.}}
\end{itemize}
We anticipate this study will benefit both the visualization and geospatial analytics communities by highlighting and clarifying the role of geospatial subject matter in recent VIS Conference publications. 

\section{Related Work}\label{relatedwork}

Inspired by claims that 80\% of all human-generated data is geospatial~\cite{Mac:2001}, prior work has sought to identify metrics to quantify the amount of geospatial data in representative data collections. Hahmann et al.\@ proposed analyzing the network of links between data on the Semantic Web to determine the degree of geospatial reference for each node~\cite{Hah:2011}. Later work from Hahmann et al.\@ used network analysis on Wikipedia article links and cognitive analysis to identify the articles as geospatial or non-geospatial~\cite{Hah:2013}. Kienreich et al.\@ proposed a geographic browsing system to anchor encyclopedia articles based on geospatial references made within the article content~\cite{Kie:2006}. For this paper, we implemented our own cognitive analysis to categorize papers containing geospatial versus non-geospatial content.


As we considered approaches to assess the geospatial content of IEEE VIS, we encountered a number of meta-studies focused on popular visualization techniques, such as multiple-view layouts, trees, and glyphs~\cite{Alm:2019}~\cite{Jur:2010}~\cite{Sch:2011}~\cite{Gra:2010}~\cite{Fuc:2016}. 
Other \textcolor{black}{meta-studies} were more general. van Ham used hierarchical clusters labeled with paper keywords in order to examine discipline variability of visualization studies~\cite{van:2004}. Isenberg et al.\@ used author defined and expert chosen keywords to create hierarchical clusters of IEEE publications to identify trends and common themes in the visualization community~\cite{Ise:2014}. Isenberg et al.\@ later expanded upon this by creating a dataset to better understand trends of research in the visualization community~\cite{Ise:2017}. This work produced a VisPubData meta-collection which has continued to be updated since its publication in 2017 and includes paper titles, abstracts, authors, DOI, and other metadata for each paper since 1990. In our work, we used elements of this dataset to situate our findings about recent papers within a longer-term context.  


\section{Identifying Geospatial Analysis}\label{methodology} 
Given our core mission to identify geospatial papers within IEEE VIS Conference papers, defining the terms ``geospatial" and ``geospatial analysis" was a critical first step. 
The latter term, ``geospatial analysis," was designated to mark the intrinsically geospatial quality of a paper. We realized that evaluating papers effectively would require establishing working definitions for both terms. 
Significant thought was given to how we might best define these terms, both because of their centrality to this paper but also for reproducibility and to ensure that all contributors used the same working definitions for the analysis. Ultimately, the research team settled on the following working definitions:
\begin{itemize}[noitemsep]
    \item{
    \textit{Geospatial}: Involves georeferenced, GPS, or satellite data that captures surface or atmospheric attributes of a planetary body. AND/OR, involves the use of a GIS or analytical techniques specifically associated with processing geographic data.
    }
    \item{
    \textit{Geospatial Analysis}: A tool or analysis that was specifically designed for geospatial data or applications. AND/OR, an analysis in which a geospatial component was fundamental to understanding the results (i.e., if you removed it, the conclusion(s) of the paper could be different). For our purposes, geospatial analysis was used as a short-hand to describe the fundamental geospatial quality of a paper.
    }
\end{itemize}
Using these working definitions, our intent was to take a broad approach to what could be considered geospatial while still limiting the assignment of the term geospatial to topics concerned with geographic relationships.

\subsection{Metadata Collection}

As an initial foray into this meta-analytical work, we decided to focus our attention on recent developments in the geovisualization space. As such, we cataloged IEEE VIS Conference publications from conference years 2017 to 2019 and collected metadata for each publication. Workshop papers were not included. For each paper, we collected the conference year, authors, title, VIS track, and session title from Open Access VIS~\cite{Har:2018}, a collection of open access papers from VIS conferences (2017-present). Papers that were listed but not linked within Open Access VIS were added manually to our metadata catalog. In total, we identified 585 papers for review. 

A detailed systematic review of each of the 585 papers was infeasible for the scope of this work. As such, we needed to select a germane subset of the papers. To maximize the number of geospatial papers reviewed, we performed a word frequency analysis for each paper, identified terms of interest, and then used those terms to assess likelihood that a given paper might be geospatial.

\subsection{Filter by Word Search}\label{filtering}

To generate word frequencies for each paper, we performed some preprocessing on the 585 documents in our corpus (PDF to text conversion, lowercasing and removing punctuation, numbers, URLs, stop words, short words, and excessively long words). 
We then created an $n$ x $m$ word frequency matrix, where $n$ is the number of unique words from the entire corpus and $m = 585$, the paper count. 

Once word counts were organized by paper, we used R to explore which words might be most useful for identifying geospatial papers. Although we initially considered automating the identification of terms of interest, we ultimately decided to manually identify words on a case-by-case basis to avoid missing relevant terms. Specifically, we combed the data using words and prefixes commonly associated with geospatial content (e.g., carto-, geo-, lati-, longi-) and then manually marked terms for inclusion in our search term list. We included words that we thought would primarily be used in a geospatial context. That said, a few general terms were included \textcolor{black}{--} such as ``map" and ``spatial" \textcolor{black}{--} in order to remain robust to authors' language choices. In total, 264 search terms were used to subset the data.

To perform the subsetting, we used the word frequencies as a scoring mechanism to identify papers more likely to be geospatial. In this way, we selected the top 220 papers with the greatest number of search term occurrences from the 585 papers we cataloged. The 220 paper cutoff (38\%) was informed by our resource constraints. Given the large number of papers that were ultimately determined to be non-geospatial after close reading, this subsetting likely captured the majority of intrinsically geospatial papers in the dataset (defined herein as papers qualifying as geospatial analysis).

\subsection{Assessing Geospatial Qualities}

To assess the geospatial qualities of the 220 papers selected, we developed questions to evaluate elements that we might expect from a geospatial paper. The questions we used and their associated column headings in our dataset (in brackets) are listed below:
\textcolor{black}{
\begin{enumerate}[noitemsep]
    \item{
    Is the title geospatial? [Geospatial Title]}
    \item{
    How many geospatial figures are included? [Geospatial Figure Count]}
    \item{
    How many total figures are included? [Total Figure Count]}
    \item{
    Does the paper use geospatial data? [Geospatial Data]}
    \item{
    Was the tool and/or analysis a geospatial analysis? [Geospatial Analysis]}
\end{enumerate}
}
\textcolor{black}{While the last two questions are self-evidently central to our research, the first three questions served as a means of exploring if the title and percent of geospatial figures could be used as an effective filter for geospatial content.}

To answer these \textcolor{black}{five} questions, we performed a blind, double-entry review. In a first round, each team member answered questions 1-5 for a distinct set of 20 papers. The second round repeated this procedure with new paper-reviewer pairings, and reviewers were not provided access to round one decisions. A final round of reviews was performed to resolve disagreements across rounds one and two. This round only re-examined conflicting responses. Responses that did not exhibit disagreement were assumed correct and excluded from this final review. Previous decisions and comments were provided to inform final review decisions.   
Final reviews were conducted in pairs so that the final answer would represent a joint conclusion between two reviewers. 

\subsection{Categorizing Papers by \textcolor{black}{Data Domain}}

After completion of this initial meta-analysis, papers identified as containing either geospatial data \textcolor{black}{(Q4)} or a geospatial analysis (Q5) were categorized by \textcolor{black}{geospatial data usage} to examine which geospatial domains had received attention in recent VIS Conference publications. 
\textcolor{black}{To this end, two reviewers examined each paper to identify where geospatial data was used. Specifically, they generated keywords and short descriptions of data used within the paper and then synthesized these observations to identify data domains and develop an initial categorization schema to describe data use in each paper.  A separate review team then coalesced this information and made final decisions on category selection.} 

\subsection{Supporting Analysis (1990-2019)}
To provide context for our core work, we also wanted to look for artifacts of geospatial data and geovisualization in IEEE VIS over time since its inception in 1990. The VisPubData meta-collection by Isenberg et al.\@ \cite{Ise:2017} currently includes metadata for 1990-2018, plus tentative metadata for 2019. Assuming the contents of the titles, abstracts, and author keyword lists are a reasonable indicator of geospatial content, we decided to leverage this meta-collection for further contextualization, adding any missing titles and abstracts, and using the first paragraph of the introduction as a proxy for the abstract when papers themselves did not provide an abstract. Then, we devised a short list of 66 geospatial key terms based on the expertise of the collaborators.  This list was more conservative than the one used to filter papers for inspection in Section~\ref{filtering} because this was to be a one-pass, automated process not followed by a close reading.  Overloaded terms -- for example: ``map" -- were excluded from this list, and terms unlikely to be used in non-geospatial contexts -- such as ``choropleth" -- were included to reduce inclusion of false positives. After prepossessing (lower-casing, tokenizing, and stemming), the title, abstract, and author keyword list of each paper (1990-2019) were searched for occurrences of each term. 

\section{Results}\label{results}
Within the top 220 papers that we investigated, \textcolor{black}{94} contained geospatial data, and\textcolor{black}{, of those 94 papers,} 64 constituted a geospatial analysis (full meta-data results are available online).~\footnote{\textcolor{bio-indigo}{\href{https://go.ncsu.edu/ieee_geovisualization_review}{https://go.ncsu.edu/ieee\_geovisualization\_review}}} Across all 220 papers, \textcolor{black}{the average percentage of geospatial figures per paper was 17.65\%.}
Of those papers that constituted a geospatial analysis, \textcolor{black}{the average percentage of geospatial figures per paper was 47.40\%.}
Only two papers had geospatial content in every figure~\cite{Arl:2019}~\cite{Zho:2018}. \textcolor{black}{For papers not classified as geospatial analysis, the average percentage of geospatial figures per paper was only 5.44\%.}

\begin{figure}[htb]
 \centering
\includegraphics[width=3.2in]{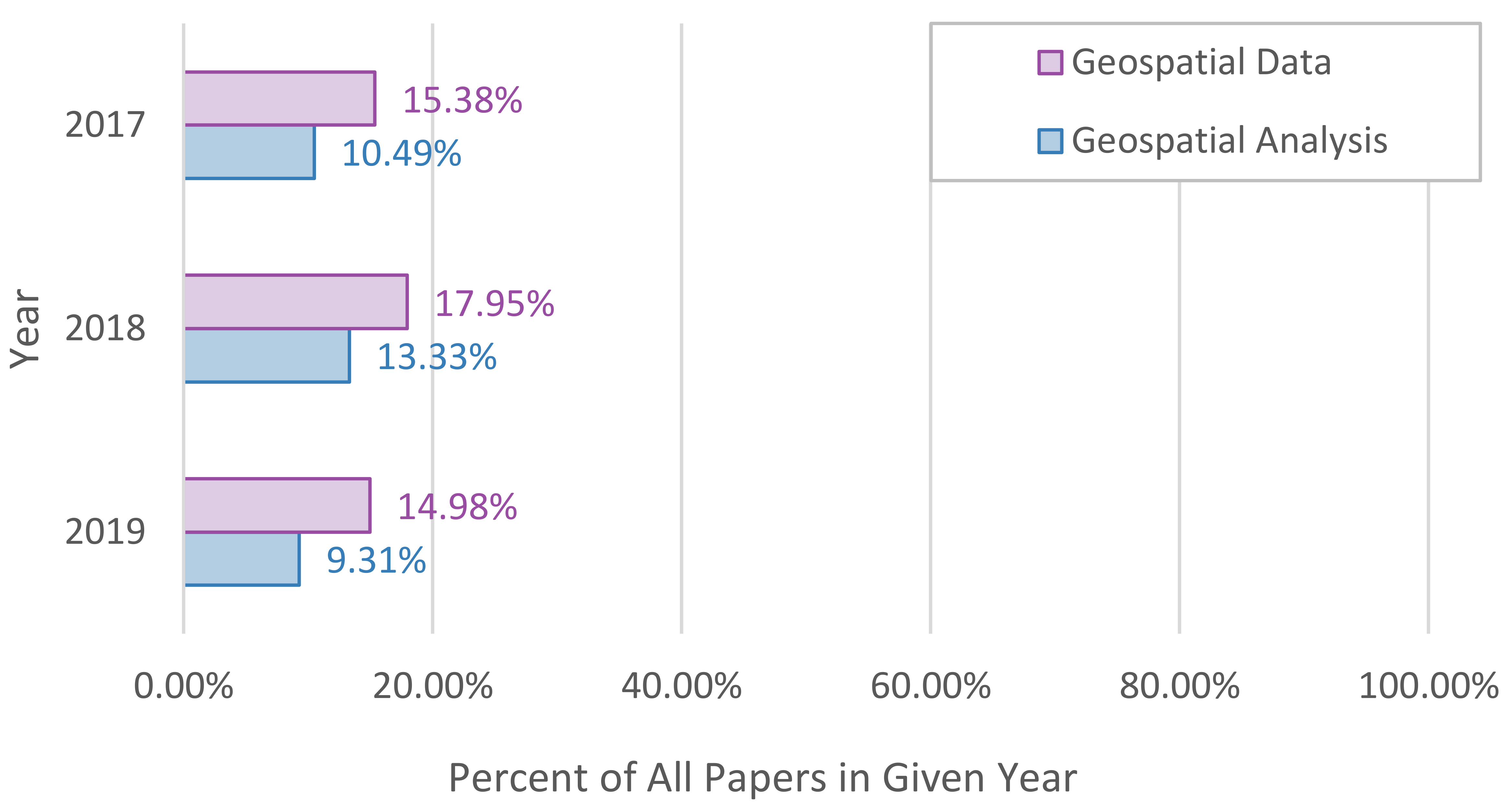}
\vspace{-.1in}
 \caption{Percentages of all VIS papers each year that used geospatial data (\textcolor{black}{purple}) or qualified as geospatial analysis (\textcolor{black}{blue}).}
 \label{fig:percentGeospatial}
\end{figure}

By year, the 2017 VIS Conference contained 15 papers that qualified as geospatial analyses, the 2018 VIS Conference contained 26 papers that qualified as geospatial analyses, and the 2019 VIS Conference contained 23 papers that qualified as geospatial \textcolor{black}{analyses} (Fig.~\ref{fig:percentGeospatial}). Because the amount of papers captured in the top 220 varied across years (2017 comprised 22.73\% of reviewed papers, 2018 comprised 39.55\%, and 2019 comprised 37.73\%), Fig.~\ref{fig:percentGeospatial} was organized to display percentages as a proportion of all papers in a given year (including those not reviewed). This decision was also informed by our confidence that we likely captured a large majority of the geospatial-related papers across all years.

\begin{figure}[htb]
 \centering
\includegraphics[width=\columnwidth]{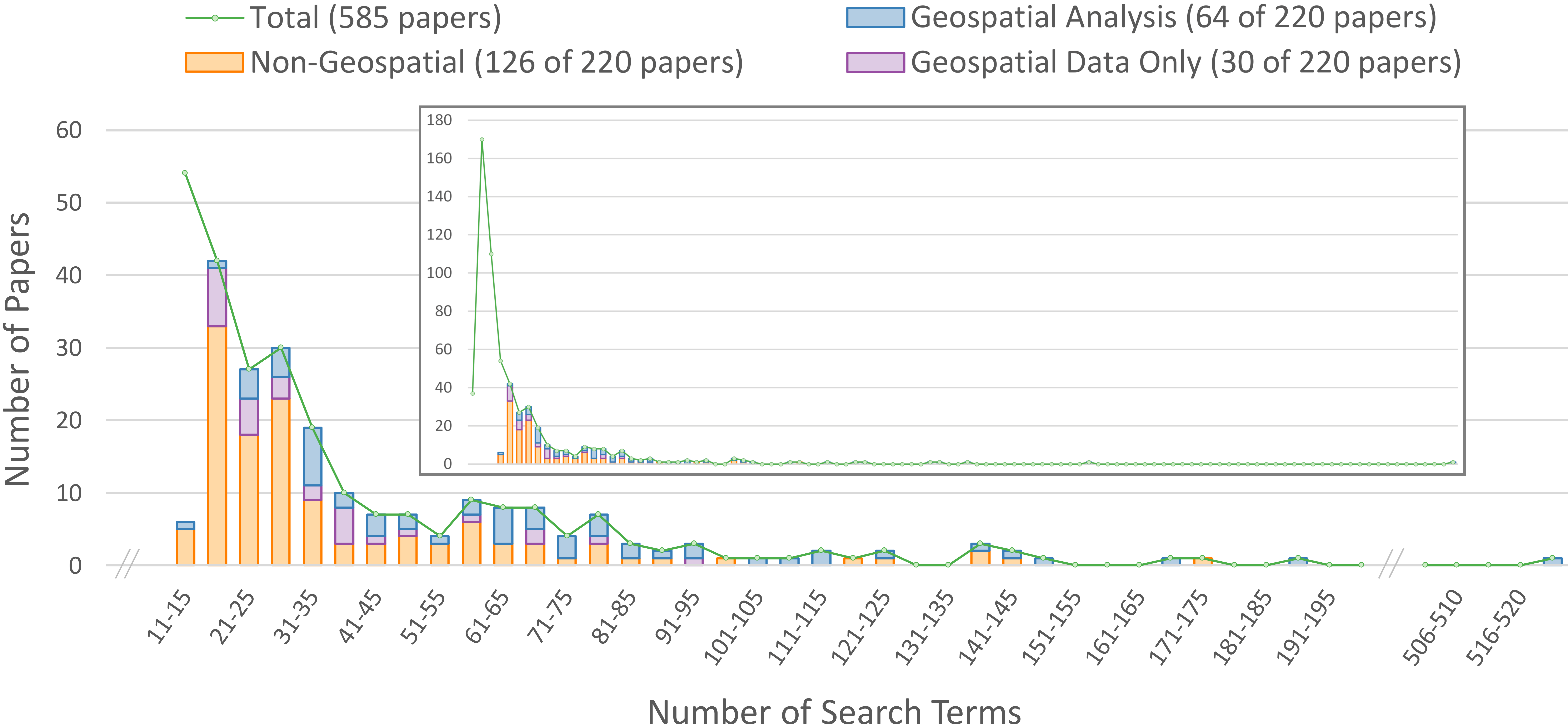}
 \vspace{-.1in}
 \caption{\textcolor{black}{Comparing search term counts to our results. The green line shows paper counts for the full 585 papers, binned by search term counts.  The bars show our results for the top 220 papers (orange, purple, and blue for non-geospatial, geospatial data only, and geospatial analysis).} The main chart has breaks near each end of the horizontal axis. The inset shows the full graph.}

 \label{fig:wordCount}
\end{figure}

In reviewing our search term frequency filtering, we observed that 33 of the 64 papers that were determined to include geospatial analysis contained 65 or fewer search term hits (Fig.~\ref{fig:wordCount}). The other 31 papers contained between 66 and 523 search terms hits. The paper with the largest amount of search term hits -- 523 -- contained 271 occurrences of the word ``cartograms" 
across its 14-page text~\cite{Nus:2018}.

\begin{figure}[hb]
 \centering
\includegraphics[width=\columnwidth]{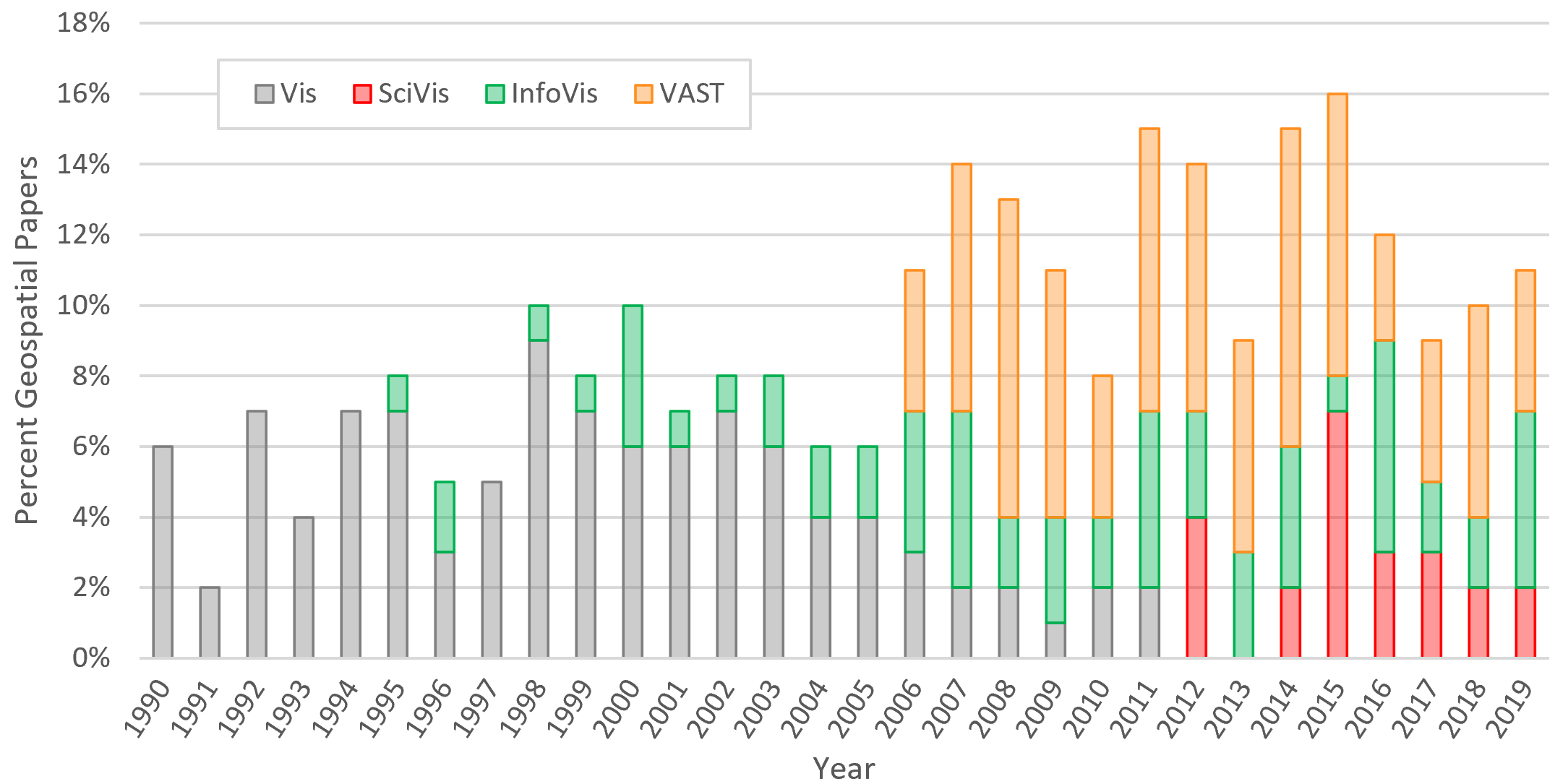}
\vspace{-.25in}
 \caption{Percentage of papers per given year from 1990 to 2019 with one or more geospatial key \textcolor{black}{terms} in the title, abstract, and/or author keyword list, arranged by conference track.}
 \label{fig:stacked_9019}
\end{figure}

As indicated by the supporting analysis, the occurrence of papers containing at least one geospatial keyword appeared to increase through time (Fig.~\ref{fig:stacked_9019}). When aggregated by decade, this pattern more clearly constituted an upward trend (geospatial metadata: 6.5\% in the 1990-1999 papers, 9.5\% in the 2000-2009 papers, and 11.9\% in 2010-2019 papers).  VisPubData contained metadata only for SciVis, InfoVis, and VAST papers; however, the OAVis collection also included journal and short papers, complicating direct comparison between these two for years 2017-2019. VisPubData contained \textcolor{black}{215} fewer papers for these years.  The supporting analysis found key terms in the metadata of 11\% of the 2017-2019 papers. Considering our close-reading decisions to hold geospatial analysis as ground truth, these 2017-2019 supporting analysis results contained \textcolor{black}{13} 
false-negatives, 15 false-positives, 20 true-positives, and 86 true-negatives, plus \textcolor{black}{236} 
negative results for papers not chosen for close reading.  

The classification of geospatial-data-using papers by \textcolor{black}{domains} yielded \textcolor{black}{14}  categories (Fig.~\ref{fig:teaser}): \textcolor{black}{(1) Multi-Domain, (2) Atmospheric Science, (3) Movement, (4) Cartography, (5) Social Media, (6) Urban Planning, (7) Marine Science, (8) Demography, (9) Planetary Science, (10) Geoscience, (11) Economics, (12) Education, (13) Text, and (14) Art.} 

\textcolor{black}{Multi-Domain papers (23\%) applied their work to geospatial data from two or more domains (e.g., Li et al.~\cite{Li:2019}  visualized both air pollutants and socioeconomic data). Atmospheric Science (22\%) papers were related to climate~\cite{Kappe:2019}, air pollution~\cite{Deng:2020}, and meteorology~\cite{Kern:2019}. Movement papers (18\%) included applications using trajectory~\cite{Andrienko:2018} and origin-destination data~\cite{Andrienko:2017:pat}. The Cartography category (10\%) papers applied tools/analysis to map features~\cite{Nusrat:2018}~\cite{Reckziegel:2019}. That is, the map features were themselves considered to be the data. There were also several Social Media (9\%)~\cite{Li:2020}~\cite{Liu:2020} and Urban Planning applications (5\%)~\cite{Mir:2018}~\cite{Kau:2018}. Marine Science (4\%) chiefly captured flow visualization papers~\cite{Friederici:2018}. Smaller categories were folded into the ``Other" class in Fig.~\ref{fig:teaser}: two papers using demographic data ~\cite{Goc:2019}~\cite{Wilkinson:2018} and one each of the Planetary Science~\cite{Bladin:2018}, Geoscience~\cite{Hyde:2018}, Economics~\cite{Arleo:2019}, Education~\cite{He:2018}, Text~\cite{Chung:2017}, and Art~\cite{Crissaff:2018} domains.}

\section{The Role of Geospatial Subject Matter}\label{discussion}
Assuming our filtering captured the majority of geospatial-related papers in the dataset, 
it appeared that the presence of geospatial analysis and the use of geospatial data in IEEE VIS Conference publications remained relatively stable across 2017, 2018, and 2019 (Fig.~\ref{fig:percentGeospatial}). As we might expect, fundamentally geospatial papers were found to contain \textcolor{black}{a notably higher percentage of} geospatial figures than non-geospatial papers \textcolor{black}{on average}.


Unpacking the role of geospatial subject matter in our study proved fairly difficult. Even with concise working definitions, many decisions still required nuance, and we found that reviewers could often reasonably arrive at divergent conclusions. Additionally, we found that adhering strongly to our working definitions sometimes meant that a paper could use only geospatial data examples but still not be marked as a geospatial analysis. For example, in a paper from Lui et al.~\cite{Liu:2019}, the authors demonstrated their uncertainty visualization technique strictly on cyclone data, but, because their techniques were not explicitly for geospatial data, the paper was ultimately not counted as a geospatial analysis. Disagreements caused by these types of issues contributed to the need for the third round of reviews.

\subsection{Variability of Geospatial Content over Time}
The supporting analysis -- spanning years 1990 to 2019 -- indicated an increasing trend across decades with a fair amount of variability between years within each decade. That said, our ability to draw definitive conclusions from our supporting work was limited as the pattern in the data could represent (1) a change in the amount of geospatial papers, (2) a change in the popularity and usage of search terms we selected, or (3) both. The core analysis was potentially more robust to this because search terms were selected through \textit{a posteriori} selection.


\subsection{Geospatial Breakdown by \textcolor{black}{Domain}}
Papers that used geospatial data in our results spanned a wide gamut of \textcolor{black}{data domains} and were present in all of the VIS tracks (Fig.~\ref{fig:teaser} \textcolor{black}{TRK} band).  The SciVis, InfoVis, and VAST tracks each made up approximately one quarter of the 220 papers isolated for review in this paper. They were represented at roughly the same levels within the \textcolor{black}{94} papers that were found to use geospatial data (\textcolor{black}{16} SciVis, \textcolor{black}{28} InfoVis, and \textcolor{black}{25} VAST papers). For years, community members have raised questions around the SciVis and InfoVis separation~\cite{Rhy:2003}. \textcolor{black}{This representation} of geospatial applications across SciVis, InfoVis, and VAST tracks seemed to support the blurring of these divisions.    

The \textcolor{black}{TRK} band in \textcolor{black}{Fig.}~\ref{fig:teaser} indicates that some tracks were more frequently associated with certain topics. These associations partially, though not entirely, aligned with our expectations. SciVis has traditionally been thought to focus primarily on physical data (volume, flow, geospatial forms, etc.). SciVis is associated with \textcolor{black}{Atmospheric} Science (longest red band), but we might expect even more of the \textcolor{black}{Atmospheric Science} applications (typically physical) to be presented in SciVis. Though InfoVis has been thought to focus on abstract, nonphysical data, InfoVis is associated with the \textcolor{black}{Multi-Domain} category (long green band). This may be due, in part, to the abundance of geospatial data, such as demographics, that can provide convenient applications to demonstrate or test the versatility of a new technique on familiar data. In fact, \textcolor{black}{close to} half of \textcolor{black}{InfoVis} papers \textcolor{black}{in the Multi-Domain category} used geospatial data but were not presenting intrinsically geospatial analyses (Fig.~\ref{fig:teaser} \textcolor{black}{GEO} band). InfoVis is also associated with \textcolor{black}{Cartography} (another long green band), perhaps because maps are an abstract representation of physical data. VAST, which focuses on interactive analytics tools, is associated with Movement (long orange band) perhaps because interactive displays are often a good solution when both geographic space and time need to be expressed.

\section{Conclusions}\label{conclusions}
In our investigation of the role of geospatial subject matter in recent IEEE VIS publications, we found that \textcolor{black}{94 of the 220 papers we reviewed made use of geospatial data, and, of those papers, 64} constituted fundamentally geospatial analyses. We organized the \textcolor{black}{94} papers containing geospatial content into categories \textcolor{black}{based on the domains of geospatial data used within the papers} and contextualized how those groupings related to VIS Conference paper types and tracks.  An inventory of geospatial term usage in 30 years of IEEE Visualization metadata indicated an increasing trend across decades.

Some questions that we would have liked to address remain unanswered. For example, though we applied close-reading to more than a third of papers from 2017-2019, a more comprehensive review could provide more information about the context of the use of geospatial subject matter and possible research gaps that exist (potential opportunities). To that end, we believe future analyses would benefit from more sophisticated text analysis methods, such as text classification, for filtering papers of interest and identifying topic groupings.


Reflecting on the research we did review, we here consider the attractiveness of geospatial data and tools for communicating research. We note that while geospatial analysis papers often presented visualization tools designed specifically for geospatial applications, a number of authors developed tools for non-geospatial purposes and included applications of those tools to geospatial tasks. These authors may be motivated not only by the abundance of multivariate geospatial data, but also by the accessibility of geospatial data for a broad audience, as geospatial data and tools are widely used across scientific domains and are also commonly consumed by the general public.




Understanding of the role of geospatial subject matter in recent IEEE VIS provided by our work can be used for several purposes. 
Though this paper focuses on observations of interest to the visualization community, the results also provide a roadmap for geospatial audiences interested in recent developments in IEEE VIS.  The results can also be useful for teaching the topic of geovisualization.   Furthermore, our work may serve as a foundation for future meta-analytical investigations across other bodies of literature for a view of how geospatial content is leveraged and thought about in the broader scientific community.




\bibliographystyle{abbrv-doi}

\bibliography{bib/combo}
\end{document}